# Behaviour of laser induced incandescence signal with long laser pulse duration and high fluence

Mario Ditaranto* and Stine Hverven

*SINTEF Energy Research, 7465 Trondheim, Norway*

*Corresponding author: mario.ditaranto@sintef.no; Tel. (+47) 47 67 08 69

## Abstract

Laser Induced Incandescence (LII) is an experimental technique that has been extensively used for in-flame measurement of soot properties. The traditional laser used for LII has been Nd:YAG laser with 5-10 ns pulse durations. The present study investigates an unexpected LII signal behaviour in the form of a secondary peak observed when using long laser pulse duration in the range of hundreds of ns. Temporal LII signals at varying fluences (up to 4.5 J/cm$^2$) and pulse durations are presented and discussed. Although an explanation based on theoretical grounds has not been found, the study shows that the phenomenon is independent of pulse duration and that its appearance correlates with laser fluences of approximately 1.1 J/cm$^2$. Artificial signal generated by the experimental procedures cannot be excluded, but if so could not be identified. One suggestion is an amplified effect of the phenomenon observed in the literature where new particles were created from the molecular cloud during high laser fluence LII. Since LII using long pulse durations are not widely used, but present a true potential in relation to sensor technology, this unexpected LII behaviour, either physically linked to the laser-soot interaction or induced by experimental flaw, should be further watched and understood.

## 1. Introduction

Laser Induced Incandescence (LII) is a powerful technique to characterize soot locally and instantaneously during its formation and oxidation in flames [1]. Since the work of Melton [2], the many physical processes occurring on the soot during the nanoseconds time scale have been considerably improved and remaining challenges in that area are described in Michelsen et al. [3]. Q-switched Nd:YAG laser with a pulse length of 7-10 ns have traditionally been used, leading to LII signal durations of 2 -3 orders of magnitudes longer depending on particle size and laser fluence. In search for lasers suitable for industrial applications of LII, Black [4] initiated studies with fiber-laser as excitation source where pulses of 192 ns and repetition rates of 30 kHz were used [4-6]. Although confirming the presence of LII signals, the temporal LII behaviour was first investigated in McCormick et al. [6] where it was noticed that vaporization appeared at higher fluence levels than in conventional 10 ns laser duration, i.e. at 0.65 J/cm$^2$ instead of the commonly accepted 0.4 J/cm$^2$ value at that wavelength.

Experimental data from our previous work [7] with pulse durations in the order of 50 – 600 ns showed that in certain cases a mechanism causes the LII signal to rebound in the form of a secondary peak rising after the signal decay has initiated. The phenomenon was observed in conditions characterized by a high laser fluence combined with long duration pulses larger than 75 ns. Such a phenomenon is not predicted by the models. There are three related cases known to the authors, the first was reported in Beyer and Greenhalgh [8] obtained in high vacuum conditions, with large black carbon particles, and occurring long after the pulse end. Furthermore, the observed increase in LII signal could only be seen on some single shot events, but not present enough to appear on the average LII signal. In the study of Witze et al. [9], a secondary peak has been vaguely reported in at fluences of 1.26 and 2.48 J/cm$^2$ with a short pulse, but no explanation could be found. The third study is reported in Goulay et al. [10, 11], where a delayed second peak appeared at high fluence and was attributed to Laser Induced Fluorescence (LIF) as its occurrence corresponded with the lifetime of LIF. The phenomenon was however observed at 532 nm excitation laser, but not at 1064 nm where LIF are unlikely to occur at such low energetic wavelengths. It is not clear whether the phenomena observed in these studies are related, but it underlines a gap in the understanding of the LII processes in conditions not commonly used.

In the study of Ditaranto et al. [7] many experimental shortcomings impaired the analysis and failed to highlight and quantify the characteristic conditions in which the rebound phenomenon and its intensity appears. This follow up study provides a better controlled set up to tackle the shortcomings of short wavelength, uncontrolled laser fluence and beam shape. Three laser pulse durations have been used, all longer than those from conventional lasers and the analysis focuses on the effect of the fluence on the LII signal to better quantify in which conditions the secondary peak LII signal occurs.

## 2. Apparatus

The burner set up used in this study is the same as that used in Ditaranto et al. [7] and is sketched in the insert of figure 1. The 5 mm diameter central nozzle has an ethylene flow bulk velocity of 21 cm/s and the air flowing in a 97 mm diameter co-flow channel has a bulk velocity of 23 cm/s. The coaxial tube (10 mm diameter) surrounding the central nozzle had no flow in that study. The flame length is 95 mm as defined by the position where soot volume fraction tends to zero and all measurements were obtained at the radial position of maximum soot volume fraction, 19 mm above the burner exit plane.

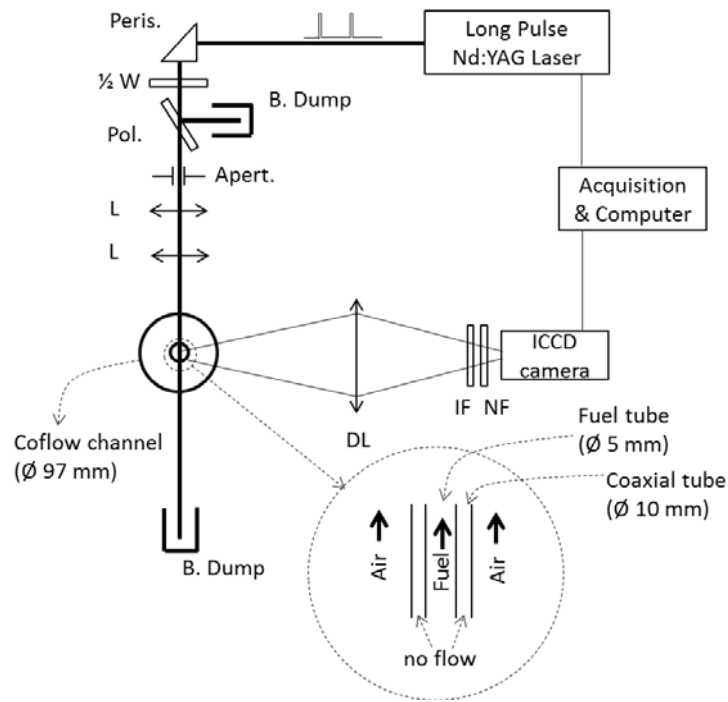

**Figure 1.** Experimental set up.

The fundamental of a pulsed Nd:YAG laser (1064 nm) with temporal shaping capabilities (AGILITE, Continuum) was focused onto the flame, following the optical path shown in figure 1. The beam was passed through a 3 mm diameter hole which was imaged in the flame by means of a 1:1 optical imaging set up. The pulse energy was controlled by a combination of a half-wave plate and a polarizing beam splitter. The intensity distribution of the laser beam has a Gaussian distribution with a 2 mm $1/e^2$ width as measured by a beam imager placed at the flame position. This diameter was used for the calculation of all fluence and irradiance values given in the remainder. To study the effect of long pulse duration and hence laser fluence and irradiance on the LII signal, pulses with near top-hat profiles of duration 100 ns, 200 ns, and 450 ns were programmed. While the energy output at the laser head was always kept constant, the beam energy through the flame was varied by rotating the half-wave plate. The pulse energy and fluence was calibrated against the angle of rotation of the plate with an energy meter placed at the flame station. The uncertainty of the laser fluence values given herein has been calculated to be better than 10 % for fluence values higher than 0.06 J/cm$^2$ based on the uncertainties of: the reading of the angle of rotation, the standard deviation measured on 200 pulses energy at each angle of rotation, and the reading of the imaged beam diameter. Experiments were performed at approximately 15 different laser fluence values for each of the three pulse durations investigated, and up to a maximum of 4.6 J/cm$^2$ for the longest duration. However only representative cases are shown in this paper for clarity.

The LII signal was imaged on the photocathode of an intensified slow scan CCD camera (Princeton PI-MAX 2) with a single 100 mm diameter lens of 250 mm focal length. A 10 nm band-pass filter centred at 488 nm was placed in front of the intensifier plate. The 26 μm$^2$ pixels of the CCD camera were 3 X 3 binned, forming a region of interest of approximately 6084 μm$^2$ since the magnification of imaging is close to unity. The time resolved LII signals were reconstructed out of several LII images obtained by gradually sliding a 5 ns intensification gate without overlap. At each time step, the signal corresponding to 200 laser shots were recorded (100 images of 2 shots accumulated on the CCD chip).

The average laser temporal profiles (shown in figure 3) were obtained by short gate imaging the Rayleigh scattering of air induced by the second harmonic of the laser (532 nm) and by sliding the intensifier gate. One hundred images of 25 laser shots accumulated on-chip were averaged at each time step after removing instances where the Rayleigh signal was overshot by parasitic Mie scattering signal of dust or particles in the air. This averaging procedure has the effect of smoothing the laser pulse time traces later shown. The laser energy output stability has been measured to have a mean fluctuation of 2 %.

## 3. Results

The time resolved LII signals obtained as a function of laser fluence are shown in figure 2 for a laser pulse duration of 100 ns. The plotted time traces have not been smoothed and show a noisy character due to the method of acquisition of the signal based on averaging the signals obtained by sliding an intensification gate and the non-linearity of the LII process. The laser output stability (2 %) induces energy variation and consequently add noise to the reconstructed temporal LII signals. Since all measurements were taken with the same acquisition properties and same position in the flame, the LII signal intensities are comparable to each other's. At low fluence the LII signal follows the typical shape corresponding to the physical processes occurring on the soot particle and extensively described in Michelsen et al. [3]. First, the soot particles having a high emissivity strongly absorb laser energy and heat up. The emitted LII signal therefore increases as temperature increases until a peak is reached, at which point the soot particles start to transfer energy out (cool) at a rate higher than the heating rate by the laser beam. The cooling mechanisms are heat transfer through conduction to the surrounding gas and radiation. The LII signal then decays at a rate dependent on the primary particle diameter. Increasing the fluence leads to an increase of the peak LII signal value that also shifts towards the start of the pulse due to higher heating rate of the soot particles. Above a certain laser fluence, the temperature reached by the particles is so high that sublimation occurs with consequent loss of mass, which becomes the dominant cooling mechanism. Further increasing the fluence accelerates the sublimation rate and the LII signal narrows in time due to a steeper decay of the LII signal.

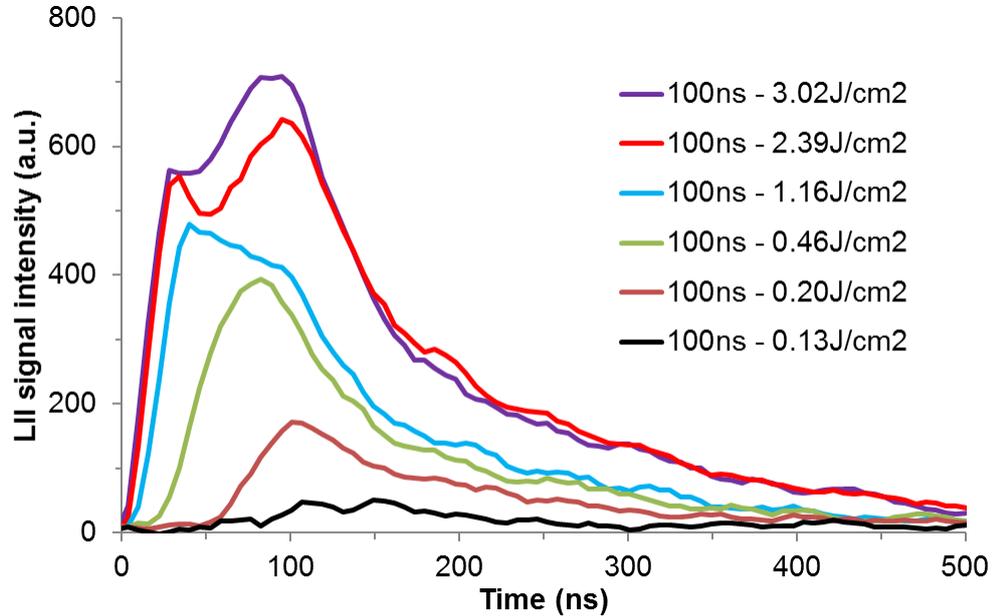

**Figure 2.** Time resolved LII signal obtained with a 100 ns laser pulse duration with varying fluence

In figure 2, for fluences higher than a value between 1.16 and 1.65 $J/cm^2$, a second peak near the end of the laser pulse is clearly noticeable confirming the findings of our previous study [7]. The secondary LII peak intensity even overshoots that of the primary peak when the fluence is greater than 2.4 $J/cm^2$. The timing of the top of secondary peak is near the position of the primary peak at low fluence incandescence. Further increasing the fluence has less and less influence on the primary peak amplitude, as expected by the models as the sublimation of soot particle limits the growth of the LII signal intensity. The maximum primary peak amplitude levels off from a fluence of 2.0 $J/cm^2$. The secondary peak maximum intensity behaves similarly as the primary peak when increasing fluence, but levels off at a somewhat higher fluence of 2.7 $J/cm^2$. The same experiments repeated with 200 ns and 450 ns pulse durations exhibit the same behaviour as seen in the representative cases at moderate and high fluences shown in figures 3(a) and 3(b) respectively. Figure 3(c) is a re-plot of figure 3(b) with the time axis normalized by the pulse duration showing that the timing of the secondary peak and general behaviour of the LII signal at high fluence is similar irrespective of pulse duration.

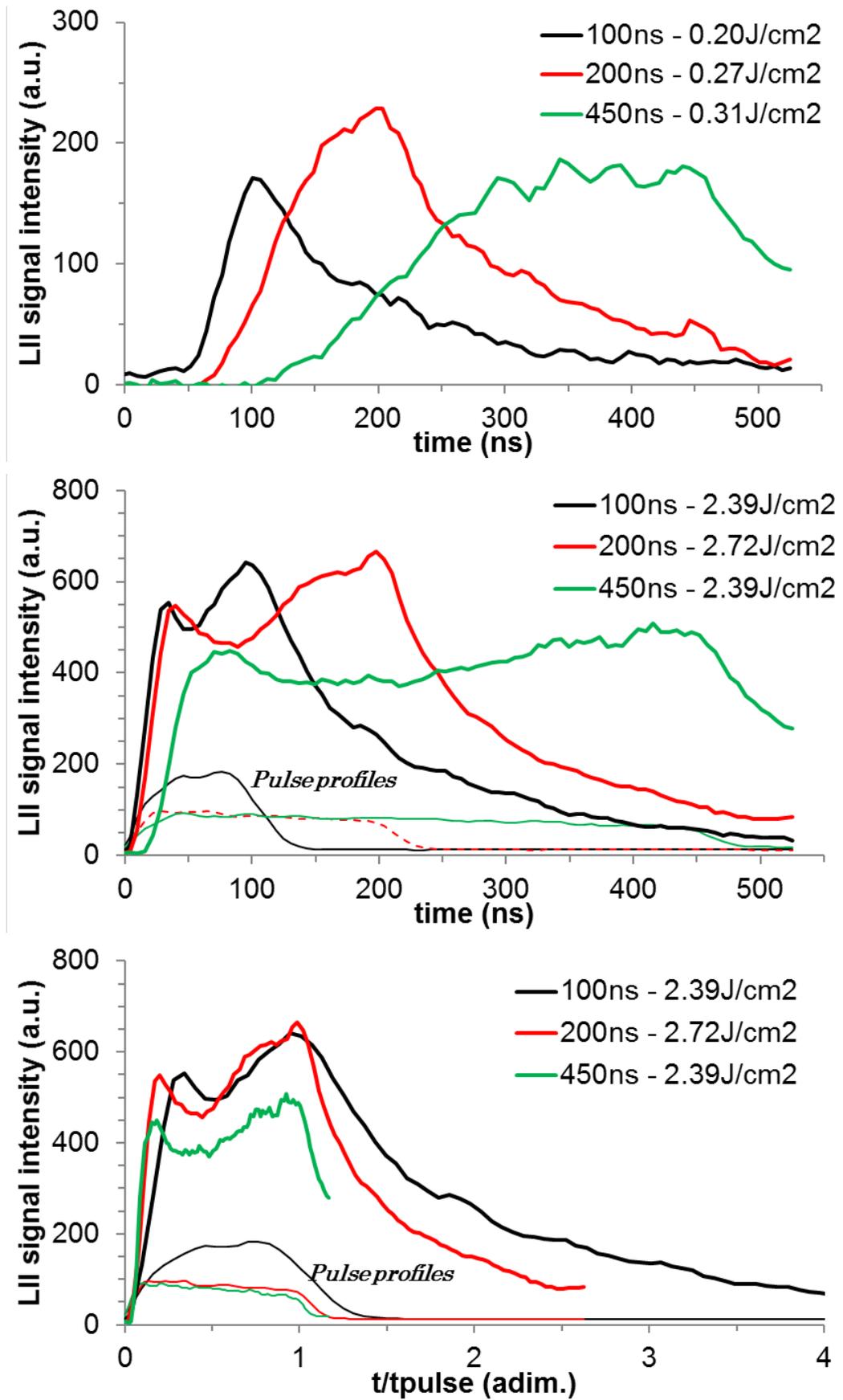

**Figure 3.** Moderate (a) and high (b) fluence time resolved LII signals obtained with laser pulse durations of 100 ns, 200 ns, and 450 ns. c) Same as b) with time axis normalized by the pulse duration. The laser pulse temporal profiles are shown as the three lower plots in b) and c)

Because of the increasing contribution of sublimation of soot particles at high fluence, the rate of LII signal increase drops with laser fluence. After a certain threshold which has been reported to be around 0.4 J/cm$^2$ and 0.2 J/cm$^2$ for laser wavelengths of 1064 nm and 532 nm respectively with pulse duration in the range 5 – 10 ns, the integrated LII signal can even flatten or decrease depending on the laser spatial profile distribution [12]. As laser fluence increases, the signal coming from the centre of the beam reaches saturation by loss of particles sooner than the signal coming from the beam wings [13]. Therefore a measurement volume hit by a laser beam with non-homogeneous spatial fluence distribution would be seen by the detector as the contribution of LII signals induced by different fluences. This effect gives rise to the so-called fluence curve plateau. A similar effect occurs if the laser has a non-homogeneous temporal pulse shape across the beam [14]. Figure 4 shows the fluence curves obtained by integrating the obtained temporal LII signal over 500 ns. As expected for a Gaussian circular beam the curves increase monotonically with fluence [12, 15], but at a rate which seems dependent on pulse durations. A clear change of slope is apparent above a fluence of 0.4 J/cm$^2$, consistent with common threshold reported in literature for traditional short pulse LII.

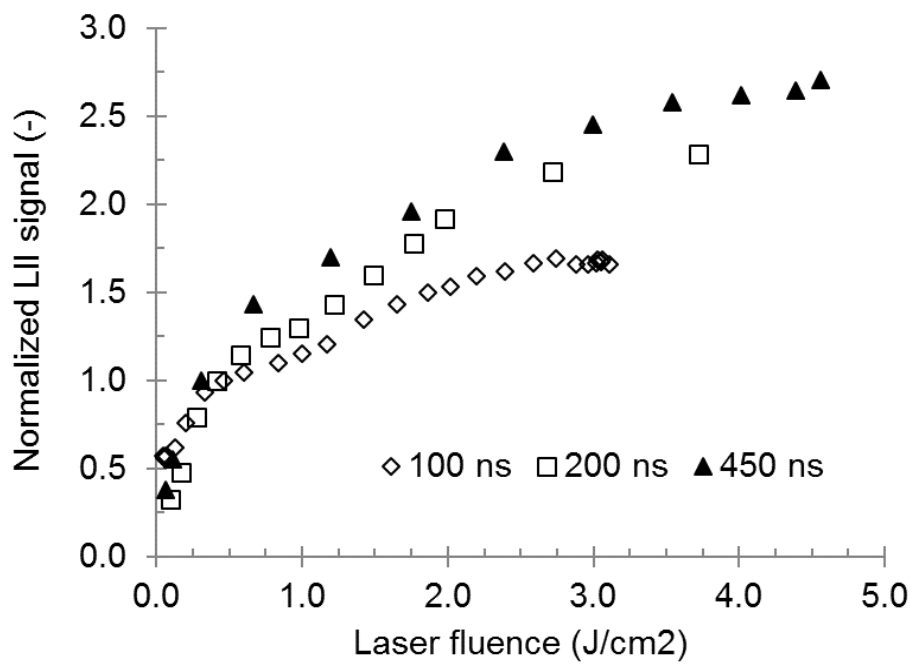

**Figure 4.**.Time integrated LII signal as a function of laser pulse duration (LII signal normalized by the value at $\approx 0.4$ J/cm$^2$)

## 4. Discussion and conclusions

The threshold values of fluence and irradiance at which the secondary peak is noticeable in the LII signals are summarized in table 1 as range (because of the discrete number of pulse energy tested). It appears that the presence of the secondary peak is not pulse duration dependent as it occurs above a fluence threshold comprised between 1.16 and 1.8 J/cm$^2$ for all cases. Irradiance, which threshold value ranges are also shown in table 1, is clearly not a representative parameter of that transition. Note that some scatter in results between the three pulse durations can be also be due the laser temporal profiles which are neither fully top hat nor totally similar in shape for all three durations (see figure 3). Another way of expressing the threshold value is by measuring the laser energy received by the soot at the time of the onset of the secondary peak at all fluences. Figure 5 shows that this energy value expressed as laser fluence and referred to as partial fluence in Witze et al. [9] is on average 1.1 +/- 0.1 J/cm$^2$ for all cases combined. In traditional 5 – 10 ns pulse duration LII, cooling by sublimation typically occurs at the same time scale as the pulse duration [16]. According to the high fluence cases of figure 2, after the LII peak, cooling by sublimation dominates heating by absorption. After a time corresponding to a partial fluence of 1.1 J/cm$^2$, a phenomenon generates a signal that suggests that absorption and consequent heating of the soot becomes again the dominant process. The actual models do not predict that behaviour which could be due to a change in structure of the soot present in the control volume. The origin could also be an experimental procedure induced error, but the modifications incurred on the set up from our previous study [7] to the present work did not affect the results. Although the temporal reconstruction methodology used induces noise on the LII signals, the appearance of the secondary peak is highly reproducible and directly correlated to high fluences, therefore not consistent with the random nature of signal noise.

**Table 1.** Onset of the secondary peak

| Pulse duration (ns) | Fluence (J/cm$^2$) | Irradiance (MW/cm$^2$) |
|---|---|---|
| 100 | ]1.16 – 1.65] | ]11.6 – 16.5] |
| 200 | ]1.49 – 1.76] | ]8.8 – 7.4] |
| 450 | ]1.20 – 1.75] | ]2.7 – 3.9] |

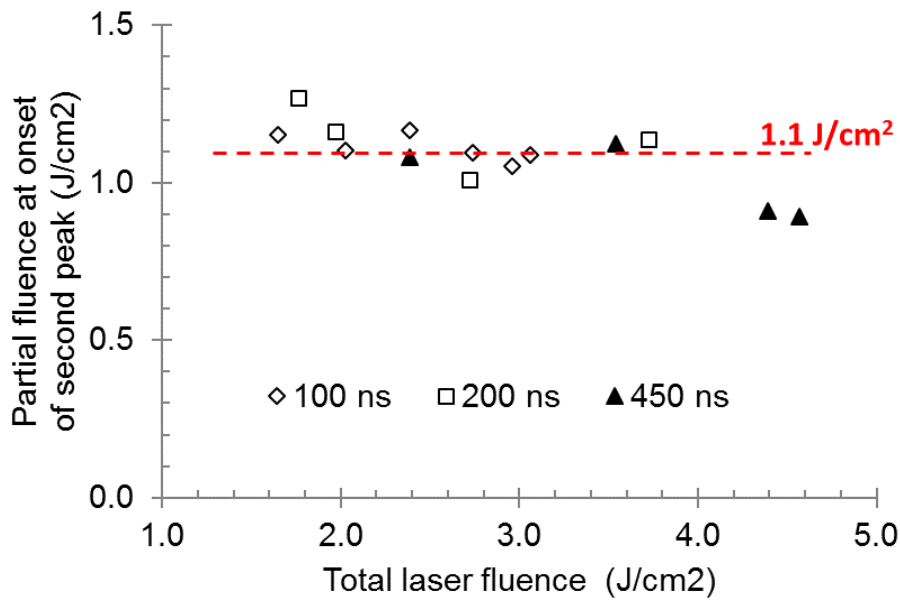

**Figure 5.** Partial fluence (laser energy received) at the onset of the secondary peak for cases exhibiting a LII signal rebound

It cannot be excluded that at these very high fluences aggregates may also disintegrate [17] and generate a discontinuity in the LII processes. Change in the absorption properties of soot during longer exposures to a laser beam has been reported by De Iuliis et al. [18] where singles particles were cumulatively exposed to several laser pulses. The absorption function E(m) was clearly changing as a consequence of graphitization, until a fluence threshold was large enough for vaporization to start and dominate the LII signal. This would occur at cumulative fluence in the range 0.8 to 1 J/cm$^2$ for soot sampled in a premixed flame, but the lower the fluence, the higher the cumulative fluence threshold to vaporization. This is obviously due to the fact that particles were allowed to cool between laser shots. Although in our experiments the fluences at which a secondary peak occur are much higher - therefore corresponding to the vaporization dominated cases in De Iuliis et al. [18] – the rate at which the energy is inferred to the particles is much slower.

Interferences from other sources than incandescence have been widely reported, particularly above the sublimation fluence where chemiluminescence and LIF can appear [19, 20]. Goulay et al. [10] evidenced spectral band emission due to excited carbon clusters as C2 and C3 at high fluence and 532 nm laser excitation followed by LIF characterized by signal decay rate with lifetime of 7 ns. However, the decay rate measured after the secondary peak have characteristics time of hundreds of nanoseconds. In Goulay et al. [20], spontaneous C2 emission centred at 467.2 nm and at 516.1 nm (i.e. near the 483 – 493 nm band collection in this study) has also been evidenced with 1064 nm excitation at high fluence. The wings of these two bands could contribute to an increase in signal, but should be minimal as shown by placing the collection spectral gate used in the C2 emission spectrum of Goulay et al. [20], and in any case not to such an extent to increase the signal to intensities higher than the main LII peak as observed in figure 2 and figure 3. Besides, the signal rise time scale is much slower than in the studies referred to above. A secondary peak has been vaguely reported in Witze et al. [9] at fluences of 1.26 and 2.48 J/cm$^2$ with a short pulse, without finding explanation, but C2 LIF effect was experimentally ruled out as the cause. In McCormick et al. [6], a LII signal is shown with a pulse duration of 192 ns and does not exhibit secondary peak behaviour. The fluence at which the measurement was made is not given in the article, but since the LII peak occurs at the very end of the pulse duration, it is probably obtained with a low fluence.

A clear observation from all the cases investigated is that the secondary peak occurs while the laser pulse is active and the final signal decay starts immediately after the pulse end. Another particular feature of the secondary LII peak is that its amplitude is as high as or higher than the primary peak, which is surprising as at these fluences the sublimation rate is high and most of the soot should have disappeared. Indeed, Yoder et al. [21] showed that the majority of sublimation occurred within the first half of the laser pulse at fluences above 0.5 J/cm$^2$ for Gaussian temporal profile. Michelsen et al. [22] showed that new particles are created out of the carbon clusters generated by sublimation and photo desorption at fluence higher than 0.22 J/cm$^2$ at 1064 nm. As shown in several studies [22-24] there particles are truly new and not a consequence of fragmentation of agglomerates. According to the study of Michelsen et al. [22], both the number density and size of the new particles increase monotonically with fluence up to at least 0.8 J/cm$^2$ (maximum fluence used in their study). They also estimated that albeit the re-distribution of particle size towards lower diameters, the volume concentration of new particles was still approximately 83 % of the initial concentration. These findings indicate that the new particles formed in the control volume as a result of LII can potentially generate significant LII signal in the same amplitude range as the first LII peak. Since the partial fluence achieved at the onset of the secondary peak is 1.1 J/cm$^2$ it can be expected that a large amount of new particles have been created, however further dedicated study is needed to support the hypothesis that the secondary signal is due to secondary incandescence of the new particles. Since LII using long pulse durations are not widely used so far, but present a true potential for sensor technology in combination with fibre lasers [5], this unexpected LII signal behaviour, either physically linked to the laser-soot interaction or induced by experimental flaw, should be further watched and understood.

## Acknowledgements

This publication has been produced with support from the BIGCCS Centre, performed under the Norwegian research program Centres for Environment-friendly Energy Research (FME). The authors acknowledge the following partners for their contributions: Gassco, Shell, Statoil, TOTAL, ENGIE and the Research Council of Norway (193816/S60).